\documentclass[fleqn,usenatbib]{mnras}

\usepackage[T1]{fontenc}

\DeclareRobustCommand{\VAN}[3]{#2}
\let\VANthebibliography\thebibliography
\def\thebibliography{\DeclareRobustCommand{\VAN}[3]{##3}\VANthebibliography}

\usepackage{graphicx}	
\usepackage{amsmath}	
\usepackage{amssymb}	
\usepackage{booktabs}
\usepackage{epsfig}
\usepackage{tablefootnote}
\usepackage{epstopdf}
\usepackage[autostyle]{csquotes}
\usepackage{longtable}

\newcommand{\asat}{{\it AstroSat~\/}}
\newcommand{\xmmn}{{\it XMM-Newton~\/}}
\newcommand{\nus}{{\it NuSTAR~\/}}
\newcommand{\swift}{{\it Swift~\/}}
\newcommand{\spitzer}{{\it Spitzer Space Telescope~\/}}
\usepackage{graphics}
\usepackage{color}

\title[X-ray study of OJ 287]{A comprehensive study of 2019-2020 flare of OJ 287  using \asat, \swift, and \nus}
\author[Prince et al.]{
Raj Prince,$^{1}$\thanks{E-mail: raj@cft.edu.pl}
Gayathri Raman,$^{2}$ Rukaiya Khatoon ,$^{3}$
Aditi Agarwal,$^{4}$ Varun,$^{5}$ 
Nayantara Gupta,$^{4}$ \newauthor
Bo\.zena Czerny,$^{1}$ 
and Pratik Majumdar,$^{6}$
\\ \\
$^{1}$Center for Theoretical Physics, Polish Academy of Sciences, Al.Lotnikow 32/46, Warsaw, Poland\\
$^{2}$ Department of Astronomy and Astrophysics, Pennsylvania State University, 525 Davey Lab, University Park, PA 16802, USA\\
$^{3}$Tezpur University, Napaam-784028, Assam, India \\
$^{4}$Raman Research Institute, Sadashivanagar, Bangalore 560080, India \\
$^{5}$ Aryabhatta Research Institute of Observational Sciences (ARIES), Manora Peak, Nainital 263002, India \\
$^{6}$Saha Institute of Nuclear Physics, HBNI, Kolkata, West Bengal, 700064, India\\
}

\date{Accepted XXX. Received YYY; in original form ZZZ}

\pubyear{2021}

\begin{document}
\label{firstpage}
\pagerange{\pageref{firstpage}--\pageref{lastpage}}
\maketitle

\begin{abstract}
OJ 287 is a well studied binary black hole system, that occasionally exhibits bright X-ray and optical flares. Here we present a detailed spectral study of its second brightest X-ray flare observed during 2019-2020 using archival \swift and \nus observations along with ToO observations from \asat.
The entire flaring period is divided into three states, defined as low, intermediate, and high state. The variation of hardness ratio (HR) with 0.3-10.0 keV integrated flux suggest a \enquote{softer-when-brighter} behavior, as also previously reported based on flux-index variations. 
Simultaneous high state X-ray spectra obtained using \swift, \nus and \asat is very steep with a power law index $>$2.
A significant spectral change is observed in
\asat-SXT and LAXPC spectrum which is consistent with \swift-XRT and \nus spectrum.
Together, optical-UV and X-ray spectrum during the high flux state, suggesting the emergence of a new high BL Lacertae (HBL) component. We have modeled the synchrotron peak with publicly available code named GAMERA for low, intermediate, and high flux state. Our modeling suggests the need of high magnetic field to explain the high state under the leptonic scenarios. Increase in the magnetic field value inside the jet could be linked to the increase in accretion rate as expected in the BH-disk impact scenario.
The color-magnitude diagram reveals a \enquote{bluer-when-brighter} spectral energy distribution chromatism during the flaring period. Different chromatism or no chromatism at various occasions suggests a complex origin of optical emission, which is believed to be produced by disc impact or through synchrotron emission in the jet. 
\end{abstract}



\section{Introduction} \label{sec:intro}
Blazar OJ 287, having the most massive black hole among all blazars, is located at a redshift 0.306 (\citealt{Mao2011}). It is classified as BL Lac object in the 3FGL (\citealt{3FGL}) \& 4FGL (\citealt{4FGL}) catalog. It is known for its high radio, optical, X-ray, and $\gamma$-ray variability. The variability time scale ranges from hours to decades across the wavebands (\citealt{Goyal_2020}). OJ 287 got more attention after the detection of high optical activity recurring at a regular interval of 12 years periodicity. In fact, this was the first AGN classified as a blazar, whose periodic behaviour was observed after analysing its century long optical light curve, which challenges the general scenario of blazars where flares or high activity states are randomly produced or in other words blazars have very stochastic variability across the electromagnetic spectrum. Periodic behavior observed in optical light curve encouraged many authors to study this source in detail and provide the possible physical explanation of periodic behavior that is new to a blazar. In literature, \citet{Sillanpaa_1988} proposed a model of binary black hole system where they associated the periodic variations to tidally induced mass inflows from accretion disk into the black hole (BH). They suggested that during the periastron passage, i.e., when the secondary black hole passes close to the primary BH, an optical outburst is produced. It has been observed that sometimes flares appear a bit earlier or a bit later than their predicted time, which is difficult to explain from first principles for a binary BH model. Later, \citet{Lehto_1996}, have performed a detailed analysis of substructure inside the major flares and they modified the binary BH model to explain the sharp flare. In their model the secondary BH crosses the accretion disk of primary BH and produces the major optical flares. Other possible models have also been proposed in the recent past. The study of \citet{Britzen_2018} considering the 15~GHz radio observation suggests the highly Doppler-boosted jet emission for major flare and the precession of jet to explain the periodicity in the light curve. Their model suggests that the optical emission is produced due to synchrotron radiation of relativistic electrons inside the jet. Many more studies suggest that the BBH model is more successful in explaining the double-peaked optical flare (\citealt{Dey_2018}).
In the Binary BH model it is shown that the impact of the secondary onto the accretion disk of the primary BH creates two hot bubbles of gas on both sides of the disk which radiate in optical through thermal Bremsstrahlung after becoming optically transparent. Since the secondary BH crosses the disk two times in each orbit, a double-peaked flare is observed/produced. The orbital period of the secondary BH is associated with the $\sim$12~years periodicity seen in the light curve.
Recent study by \citet{Dey_2018} has confirmed the presence of 24 optical flares between 1888 to 2015 by the BBH model. The 2015 flare was also explained by \citet{Dey_2018} in the context of binary black hole model.  \citet{Valtonen_2019} have estimated the values of the physical parameters of the impact of secondary onto the accretion disk of the primary BH in the BBH model.

In the past decade many broadband studies have been carried out on this source. 
In the broadband analysis, this source is classified as a low-peaked BL Lac source where the synchrotron emission is constrained by NIR to soft X-ray as the synchrotron peak in the spectral energy distribution (SED) lies below 10$^{14}$~Hz. The high energy hump of the SED generally covers the X-ray and gamma-ray frequencies, with the peak at around 0.1~GeV to few GeV (\citealt{abdo_2010}, \citealt{Kushwaha_2013}). Synchrotron self-Compton (SSC) is the most well accepted mechanism that can explain the second SED peak. However, in some flares it has been found that the SSC model is not sufficient to explain the high energy part of the spectrum and hence people have used additional external thermal field to explain this. In \citet{Kushwaha_2013}, they have used external photon field at a temperature of 250~K as a source of external seed photons. These photons get up-scattered by the highly relativistic electrons present in the jet through the inverse-Compton mechanism. 

In X-rays, this source has been studied in detail during several active states. The maximum number of observations with the longest durations have been carried out with the \swift-XRT telescope, which nicely covered many bright X-ray flares of OJ 287. From the earlier X-ray flares it has been found that OJ 287 sometimes shows a significant change in the optical-UV and X-ray spectrum and that eventually leads to the shift in synchrotron peak towards higher energy during flare. Significant change in the X-ray spectrum and emergence of new HBL component is very rare in blazars. However, this kind of behavior has been seen in a few sources studied in the past (\citealt{Pian_1998}, \citealt{Raiteri15}, \citealt{Kapanadze_2018}, \citealt{kushwaha_2018a, kushwaha_2018b, Kushwaha_2020}).

In 2020, a bright optical-UV, and X-ray flare was reported in various ATels\footnote{http://www.astronomerstelegram.org/?read=13677}(\citealt{ATel13637}, \citealt{ATel13658}, \citealt{ATel13677}, \citealt{ATel13755}, \citealt{ATel13785}). 
Recently, \citet{Komossa_2020}, \citet{ 2021POBeo.100...29K, universe7080261, Komossa2021c} presented a detailed analysis with the data from \xmmn and \nus telescopes. Many \xmmn archival observations along with recent 2018 and 2020 observations were analysed and they found that the spectrum in 2020 behaves very differently from archival observations including the year 2018. 
They suggested a non-thermal origin of the 2020 flares.
 We have performed a broadband SED modeling of the flares during 2017--2020 and it is presented in \citet{Prince-2021}.

In this work, we provide a detailed analysis of \textit{Swift}-XRT and UVOT instrument data and report the emergence of a new HBL component which eventually leads to the shift in synchrotron peak towards higher energy. The synchrotron emission is constrained by the optical-UV and X-ray data. Emergence of HBL component in low BL Lac blazars is very rare. We also carried out a Target-of-Opportunity (ToO) observation of the source using India's first space-based multi-wavelength mission, \asat (\citealt{Agrawal_2017}). The results from Soft X-ray focusing Telescope (SXT) and Large Area X-ray Proportional Counter (LAXPC) instrument of \asat are also discussed in this work. At the end we have discussed the color-magnitude diagram followed by the summary and conclusions.

\begin{figure*}
    \centering
    \includegraphics[scale=0.6]{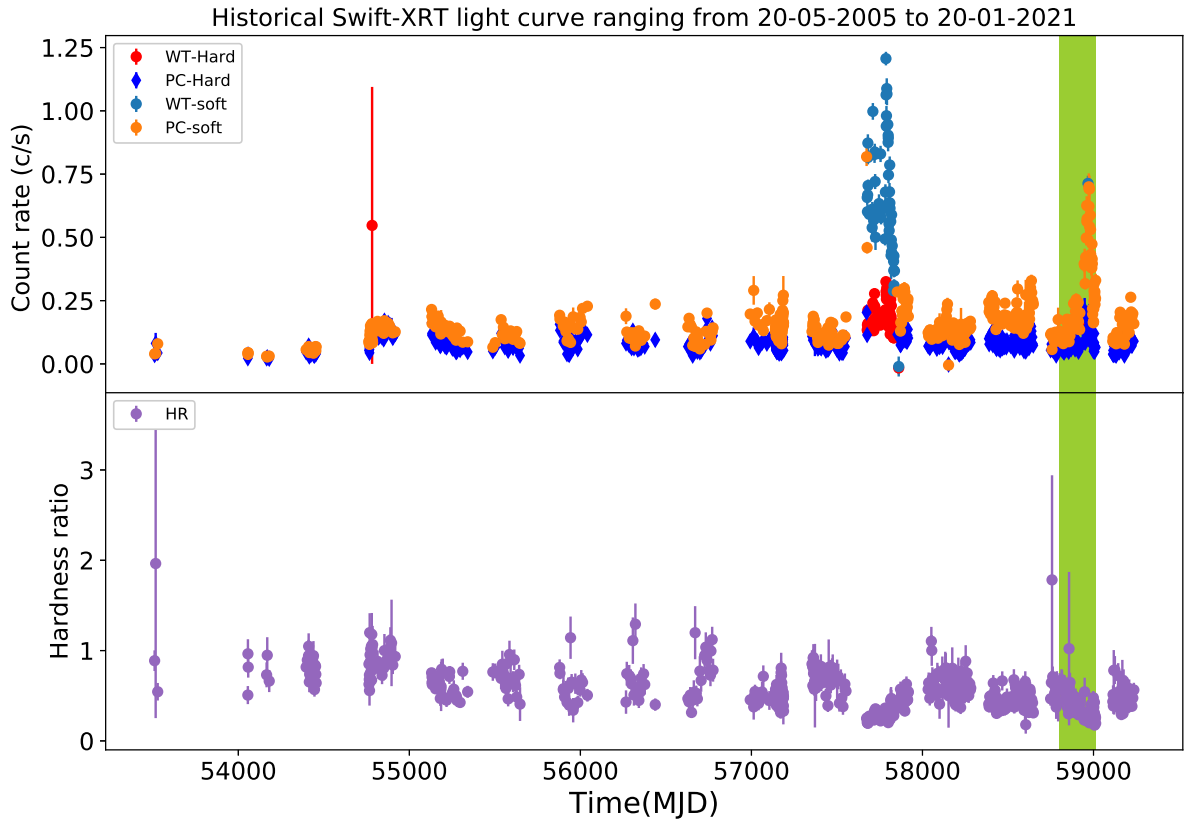}
    \caption{Long-term X-ray light curve from \swift-XRT. WT: Window timing, PC: Photon Counting, Hard: 1.5--10.0~keV, Soft: 0.3--1.5~keV.}
    \label{fig:historical}
\end{figure*}

\begin{figure*}
    \centering
    \includegraphics[scale=0.5]{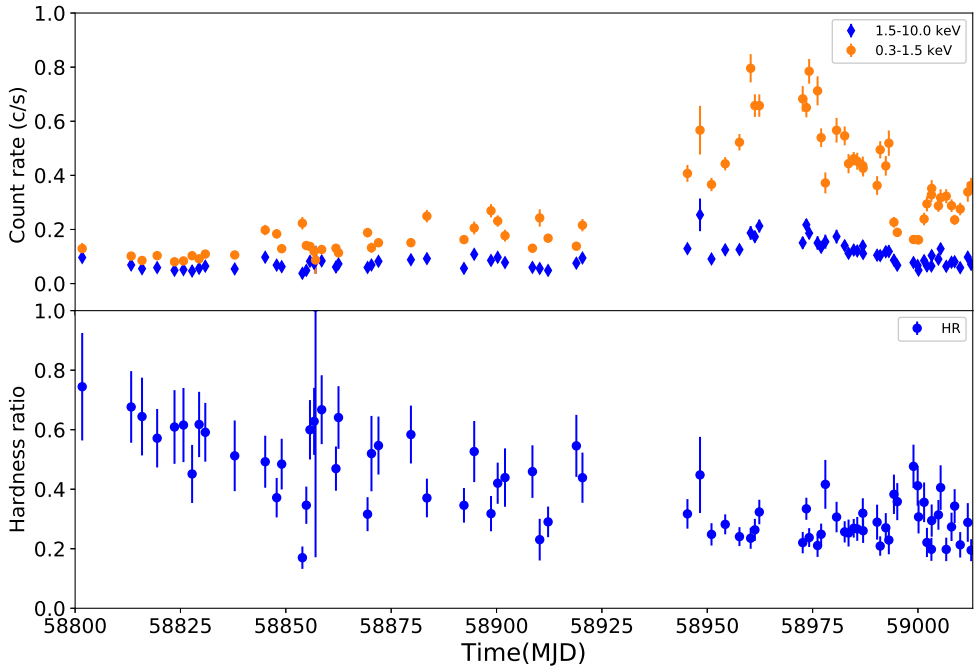}
    \caption{\swift-XRT count-rate light curve in hard (1.5--10.0~keV) and soft (0.3--1.5~keV) state is shown in the upper panel. Lower panel shows the hardness ratio.}
    \label{fig:soft-xray}
\end{figure*}

\begin{figure}
    \centering
    \includegraphics[scale=0.5]{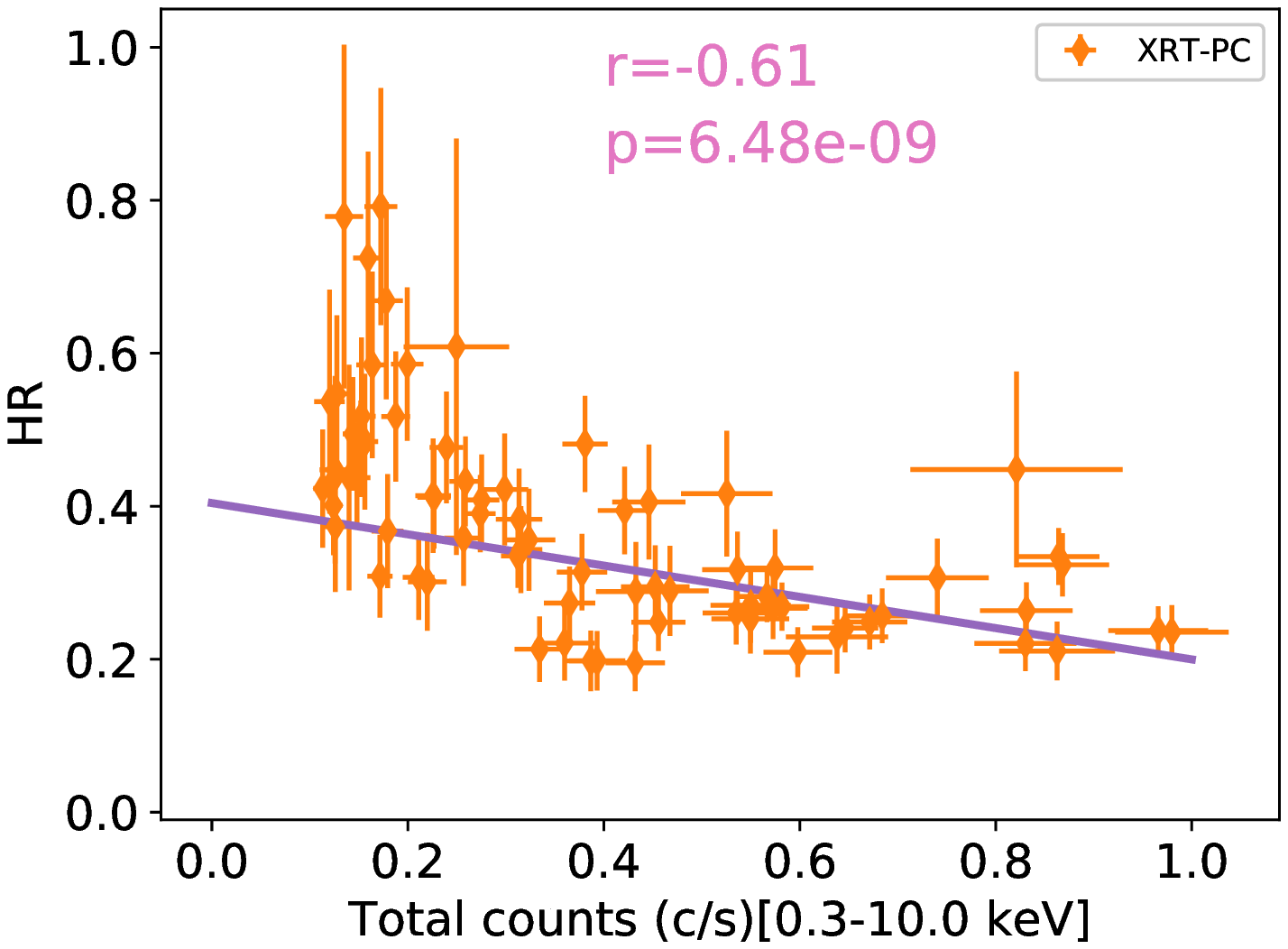}
    \caption{Hardness ratio plotted against the total count rate (0.3-10.0 keV). The $r$ and $p$  are Pearson's correlation coefficient and null-hypothesis probability. Data are fitted with straight line to show the "softer-when-brighter" trend.}
    \label{fig:HR}
\end{figure}

\section{X-ray Analysis}
\textbf{Swift-XRT:}
OJ 287 is continuously monitored by the \swift telescope under the project summarised in \citet{komossa2017}.
On February 3, 2017, an X-ray flare was observed by the \swift telescope (\citealt{Gehrels_2004}), and the results were reported in Atel 10043 (\citealt{Grupe_2017}). It is reported as the brightest flare ever detected since the monitoring of the \swift telescope. A long-term light curve is presented in \citet{Komossa_2020} where multiple flares were observed in X-rays. 
\swift-XRT is a soft X-ray telescope onboard the Swift Satellite and is sensitive in 0.3--10.0~keV energy band. 
The BL Lac OJ 287 was observed by \textit{Swift}-XRT telescope during the multiple flaring episodes between the period November 2019 to May 2020. 
We have followed the standard analysis procedure to reduce the raw \swift data as it was done in \citet{Komossa_2020}. 
The spectral fitting is done for an energy range of 0.3--10.0~keV with the Galactic absorption column density $n_H$ = 3.04$\times$10$^{20}$~cm$^{-2}$ from \citet{Dickey1990}.

\textbf{AstroSat-SXT:}
The SXT observations were carried out in the Photon Counting (PC) mode and the Level-1 data was further reduced using the \texttt{sxtpipeline1.4b} (Release Date: 2019-01-04) to generate the Level-2 data products (\citealt{Singh2016, Singh_2017}). The events from the various orbits were merged using the \texttt{SXTEVTMERGERTOOL} in Julia.
Further, we used the merged event list for temporal and spectral analysis. The merged event files were accessed by the tools \texttt{Xselect} where a source region of 10 arcsec was chosen around the source and the blank sky for the background. The response files and the background spectra are provided by the SXT-POC (Payload Operation Center) team. Finally, the grouped spectra were fitted in \texttt{Xspec} in the 0.3--10 keV band, and $\chi^2$ statistic was obtained for the best fit.

\textbf{AstroSat-LAXPC:}
LAXPC (\citealt{Yadav2016}) is designed to observe sources in a very wide range of X-rays ranging from 3 keV to 80 keV. It has three identical proportional counters (LAXPC10, LAXPC20, \& LAXPC30) onboard the \asat satellite. Data used in this work is taken from the LXP20 detector. The LAXPC30 detector is switched off due to gain instability issues. The data reduction and further processing of Level-1 were carried out using the LAXPC data analysis pipeline version 3.1\footnote{Data analysis software was obtained from http://astrosat-ssc.iucaa.in/?q=laxpcData}. The pipeline code combines data from multiple orbits and also filters out overlapping segments between each orbit. Using the standard pipeline \texttt{laxpc\_make\_event}, we generate the cleaned event file. A good time interval (GTI) window was applied during the processing using the \texttt{laxpc\_make\_stdgti} tool, in order to exclude the time intervals corresponding to the Earth occultation periods, SAA passage and standard elevation angle screening criteria. Since Blazars are observed as faint sources in LAXPC, we adopted the faint source pipeline for the spectral analysis. In order to reduce background contribution from all LAXPC layers, only the top detector layer was used for the timing and spectral analysis of the LXP20 data.

\textbf{NuSTAR:} We also searched the archive for the \nus observations.
A ToO observation in \nus (\citealt{Harrison_2013}) was performed during the high X-ray flux state on 2020-05-04 starting at 20:36:09 UTC with an effective exposure time of 29.5 ks by \citet{Komossa_2020}. The data reduction is made using the latest \nus data analysis software (NuSTARDAS) version 1.9.2 provided by \texttt{HEASOFT}. To extract the source spectrum and the background spectrum, a circular region of 20$''$ and 50$''$ were chosen around the source and away from the source, respectively.

\section{Optical and UV Observations}
Having the \swift Ultraviolet/Optical Telescope (UVOT, \citealt{Roming_2005}) on board with \swift-XRT has the advantage of getting simultaneous observations in Optical and UV bands. 
\swift-UVOT has also observed the OJ 287 in all of the available six filters U, V, B, W1, M2, and W2, simultaneously with the X-ray observations. A long-term optical-UV light curve is presented in \citet{Komossa_2020}. We have used the archival data for the period between 2019-2020, and the standard analysis procedure is followed to extract the mag and the flux as it was done in \citet{Komossa_2020}. 
We have considered the region of 5 arcsec around the source as the source region, and the region away from the source as the background region. 
The magnitudes are corrected for galactic extinction by using the reddening E(B-V) = 0.0241 from \citet{Schlafly_2011} and zero points from \citet{Breeveld_2011}. Moreover, the magnitudes are converted into flux by multiplying by the conversion factor estimated by \citet{Poole_2008} and the ratios of extinction to reddening from \citet{Giommi_2006}.

\section{Results}
\subsection{X-ray Light curves}
\textbf{XRT:} We have produced the long term X-ray light curve of OJ 287, using \textit{Swift}-XRT observations which is shown in Figure \ref{fig:historical}. The XRT observed the source in two different modes \texttt{Photon Counting (PC)} mode and \texttt{Window Timing (WT)} mode.
In Figure \ref{fig:historical}, it is observed that the source showed a very bright soft-X-ray flare in 2016-2017, which was studied in detail earlier by \citet{kushwaha_2018b}. In 2020, the second brightest soft-X-ray flare was observed as also marked by color patch in Figure \ref{fig:historical}. Our focus is to study the 2020 flare in detail and compare it with the older 2016-2017 flare.

Figure \ref{fig:soft-xray} shows the patched part of the light curve. The soft (0.3--1.5~keV) and hard (1.5--10.0~keV) count-rate light curves are shown in the top panel, and the lower panel shows the hardness ratio, which is defined as the ratio of hard counts to soft counts. From the upper panel it is clear that the blazar OJ 287 is highly variable and flaring in soft X-rays, while the count rate is lower and the source is far less variable in the hard X-ray band. In the beginning of the light curve (at MJD 58800), the counts are almost similar in both the bands, but as time passes, the soft X-ray part starts rising slowly with respect to hard X-ray. During April 2020 ($\sim$MJD 58940), the soft X-ray counts are almost four times higher than the hard X-ray counts. It remained in a very high state for an entire month till May 2020 because of which the flare was eventually recognized as the second brightest X-ray flare in the history of OJ 287. It was first recognised by the \citet{Komossa_2020}. To characterize the variability in the soft and hard X-ray band, we have estimated the fractional variability (F$_{\rm var}$) amplitude (\citealt{Vaughan_2003}). The source is found to be more variable in 0.3--1.5~keV band with F$_{\rm var}$ = 0.38$\pm$0.01 and less variable in 1.5--10~keV band with F$_{\rm var}$ = 0.27$\pm$0.01. 
In the lower panel it can be seen that, as the
source counts increases, the hardness ratio starts to shift towards the lower value. To see the clear trend in the hardness ratio (HR), we have plotted the hardness ratio against the total (0.3--10.0~keV) counts in Figure \ref{fig:HR}. The HR is anti-correlated with the total counts with Pearson's correlation coefficient 0.61, suggesting a \enquote{softer-when-brighter} trend, as also previously reported by \citet{Komossa_2020}.
This trend in X-ray is a prevalent behavior in history of OJ 287 except the flare in 2015 where the spectrum was rather harder with flux. Comparing the softer behavior detected in the current flaring state, with the harder behavior observed in the 2015 flare, we conclude that these flares may have different origins. 

\textbf{SXT:} We carried out a Target of Opportunity (ToO) observation using \asat (ObsID: 9000003672) during the X-ray flare in May 2020, since an enhanced X-ray flux was reported in several ATels. The source was observed from 15th--19th May, 2020. We have analyzed all the observations separately. In some of the observations, \asat detected a very low count rate and was heavily background dominated. Due to poor statistical quality of the data from those Obs IDs, we did not carry out any timing or spectral study for those observations. The SXT spectral products corresponding to May 15, 2020 had a good signal to noise ratio (SNR) for meaningful spectral study, which we have discussed in Section 4.4.

\textbf{LAXPC:} Along with SXT, simultaneous LAXPC observations were carried out between May 15, 2020 to May 19, 2020.(ObsID: 9000003672). We carried out a similar data reduction process for all the observational data taken during this period. As was the case with SXT, only the data of the observation corresponding to May 15 had a good SNR and is therefore presented in this paper. All the other observations were of poor statistical quality and background dominated and we do not utilize them for this study. The simultaneous observations by LAXPC with SXT provide an opportunity to model both the spectra together. The result of this joint fit is discussed in section 4.4.

\textbf{NuSTAR:} We also looked for the timing analysis in \nus, but were unable to extract any light curve. Therefore, the study is focused on understanding the spectral behavior. Detailed discussion is provided in Section 4.4.

\begin{figure*}
    \centering
    \includegraphics[scale=0.6]{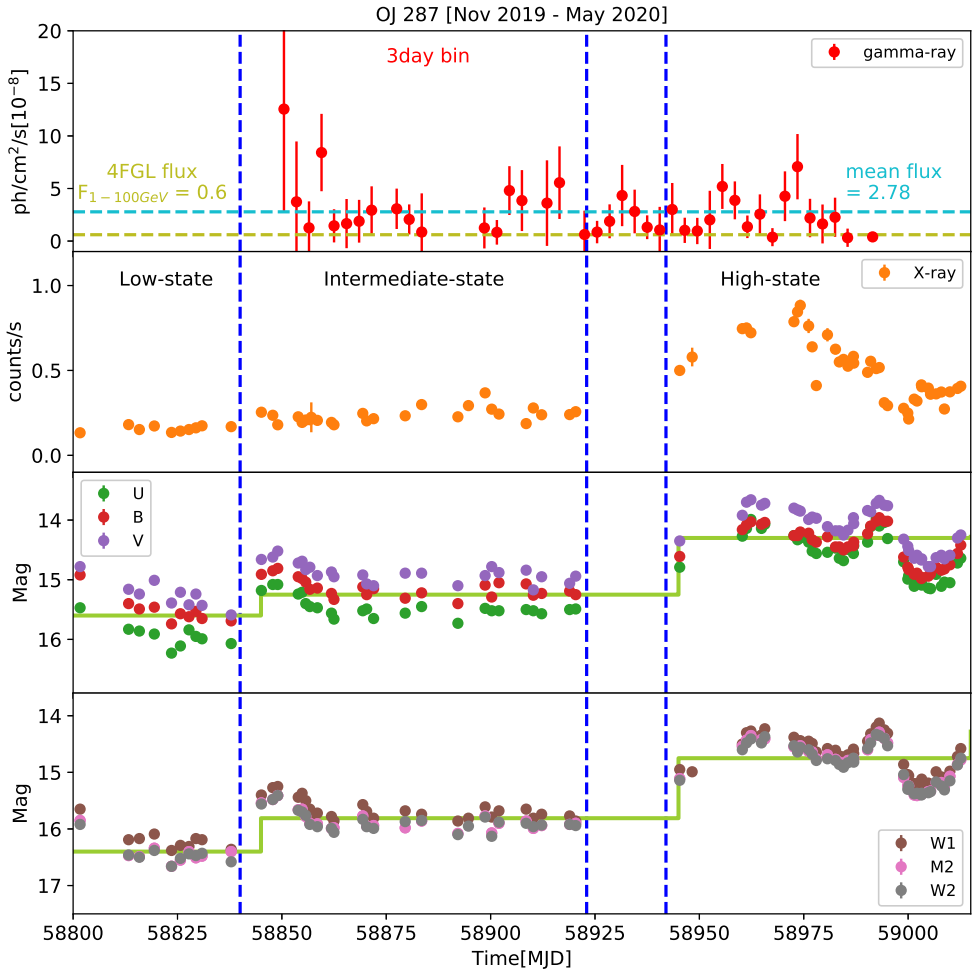}
    \caption{Broadband light curve created for period starting from November 2019 to current flaring state May 2020. Panel 1: represents the $\gamma$-ray data from Fermi-LAT; Panel 2: shows the X-ray data from \swift-XRT; Panel 3 and 4: shows the optical and UV data from \swift-UVOT. Step plot in optical-UV is shown to identify the step-wise increase in their flux.}
    \label{fig:mwl-lc}
\end{figure*}

\subsection{Optical-UV and $\gamma$-ray Light Curves}
In Figure \ref{fig:mwl-lc}, we have shown 
the simultaneous gamma-ray, X-ray , optical and UV light curves during the flaring period of the source from MJD 58800 to 59012. It can be seen that the optical-UV emissions follow the X-ray emission; \citet{Komossa_2020} show a increase in brightness with time using long-term optical and UV light curves. However, in our study, we have focused on the detailed structure of light curve during the short period between 2019-2020. Based on the average values of magnitudes (shown by a solid green line) observed in the optical-UV band, we divided the light curves into three different parts, namely, low state, intermediate state, and high state. The solid green lines which represent the average values of the magnitudes in three different states are jointly represents a step function structure. 
 In X-ray, the flux follows a similar kind of step function, but no apparent trend is seen.
 These three different states are used for the further study. No evident activity is seen in the $\gamma$-ray light curve shown in the uppermost panel. The flux observed in $\gamma$-rays is very low and mostly constant over time (fluctuating around the mean). The average flux during the flaring period is obtained as 2.78$\times$10$^{-8}$~ph cm$^{-2}$ s$^{-1}$ and shown by a horizontal dashed line. Similarly, the 4FGL catalog flux is also represented by the horizontal dashed line in Figure \ref{fig:mwl-lc} which is way below the average flux of the source during this period. Previous studies on the flares detected in 2015 and 2016-2017 by \citealt{kushwaha_2018a, kushwaha_2018b} also show no strong activity in $\gamma$-ray, though the source was flaring in optical-UV and X-rays.

\subsection{X-ray Spectral Analysis}
We have produced the X-ray spectrum of all the three states identified in Figure \ref{fig:mwl-lc}. The spectra are fitted with the simple power-law and the log parabola models, in built in \texttt{Xspec}, and a galactic absorption \texttt{tbabs} is added to the model. 
The final model we used is \texttt{tbabs*powerlaw/logparabola} and
the best fit model parameters are presented in Table \ref{tab:table1}. We also present \texttt{F-test} results to compare the different model fits and find with a p-value of $>$0.01, power-law is preferred over log parabola model.

\begin{figure}
    \centering
    \includegraphics[scale=0.36,angle=-90]{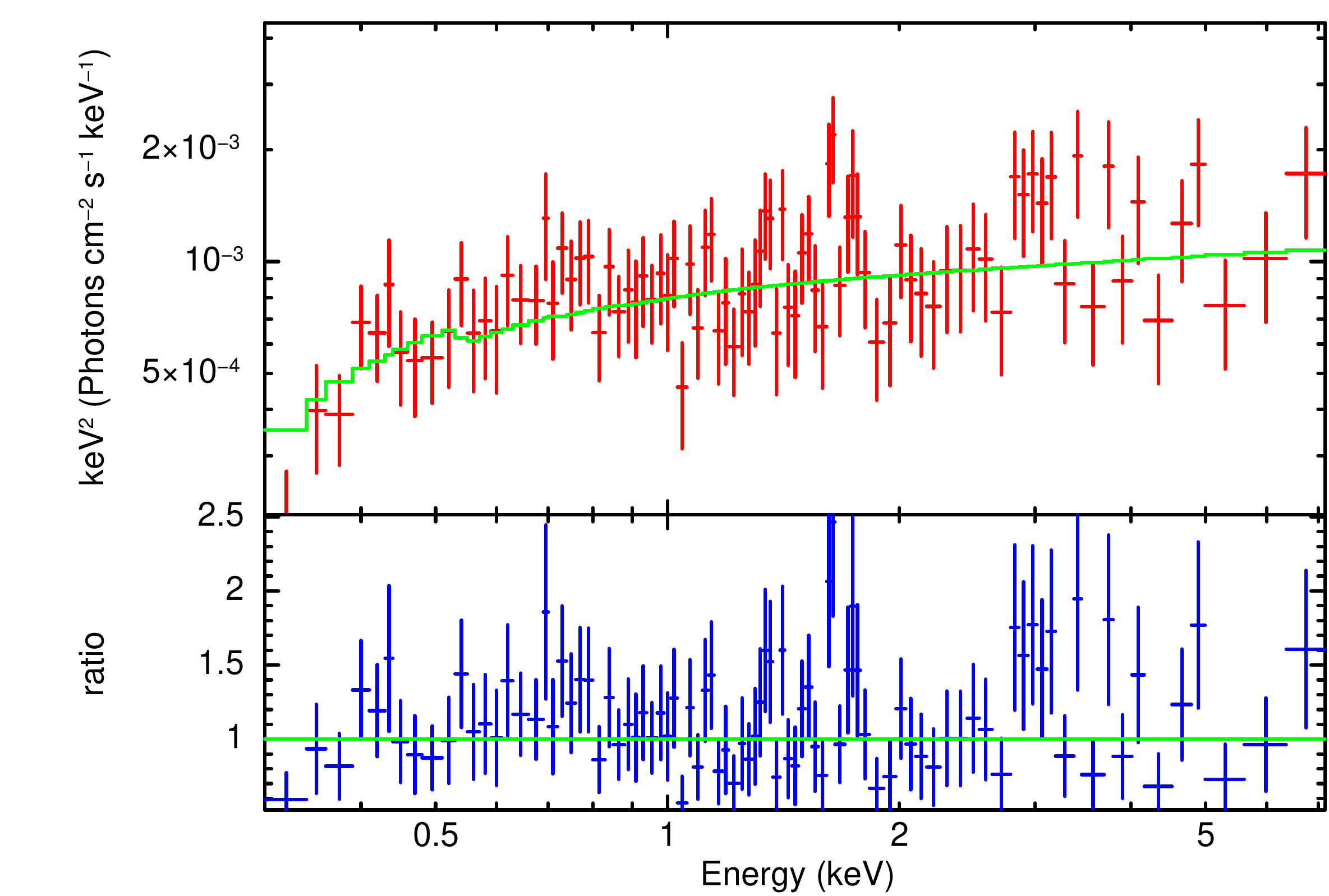}
    \includegraphics[scale=0.36,angle=-90]{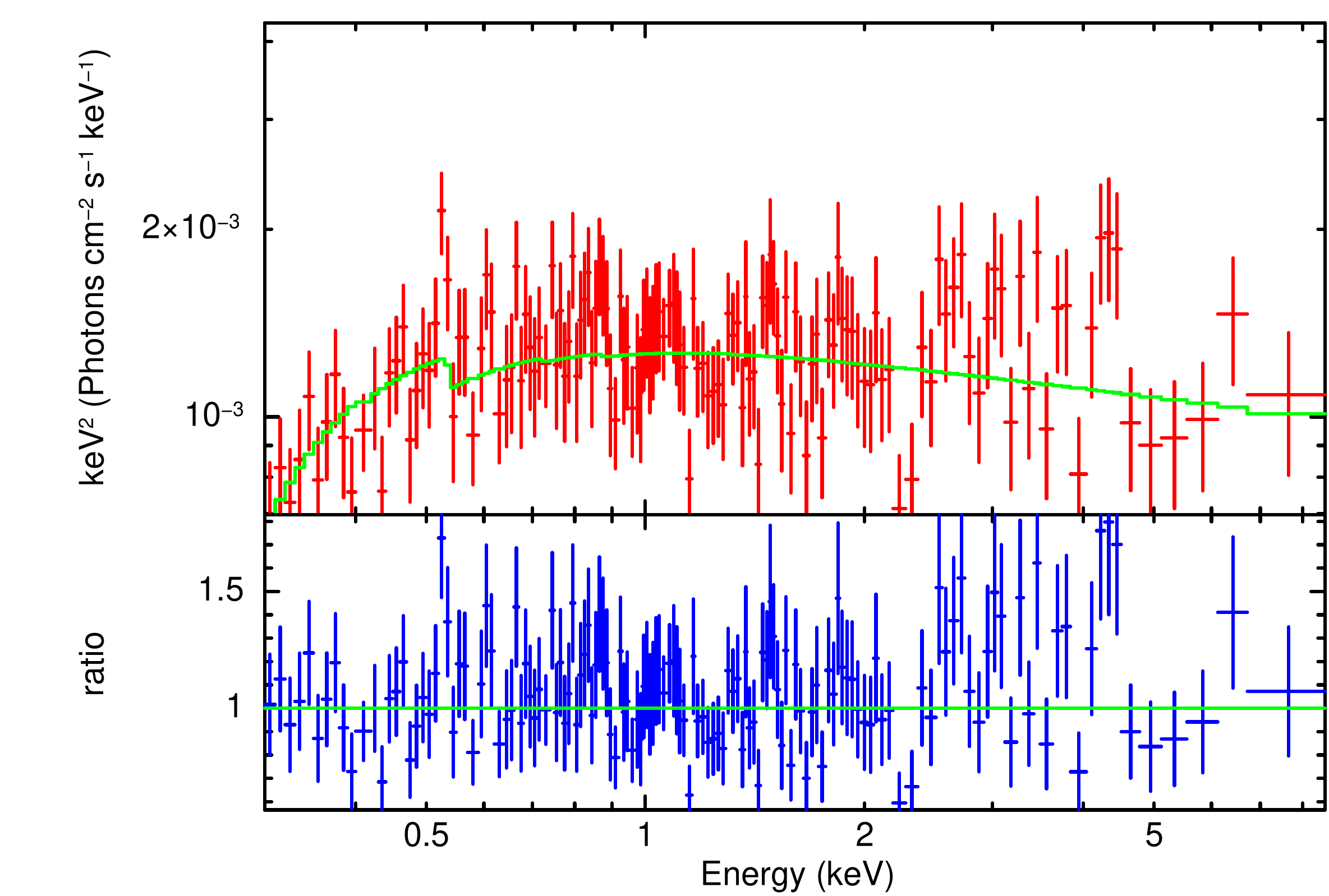}
    \includegraphics[scale=0.36,angle=-90]{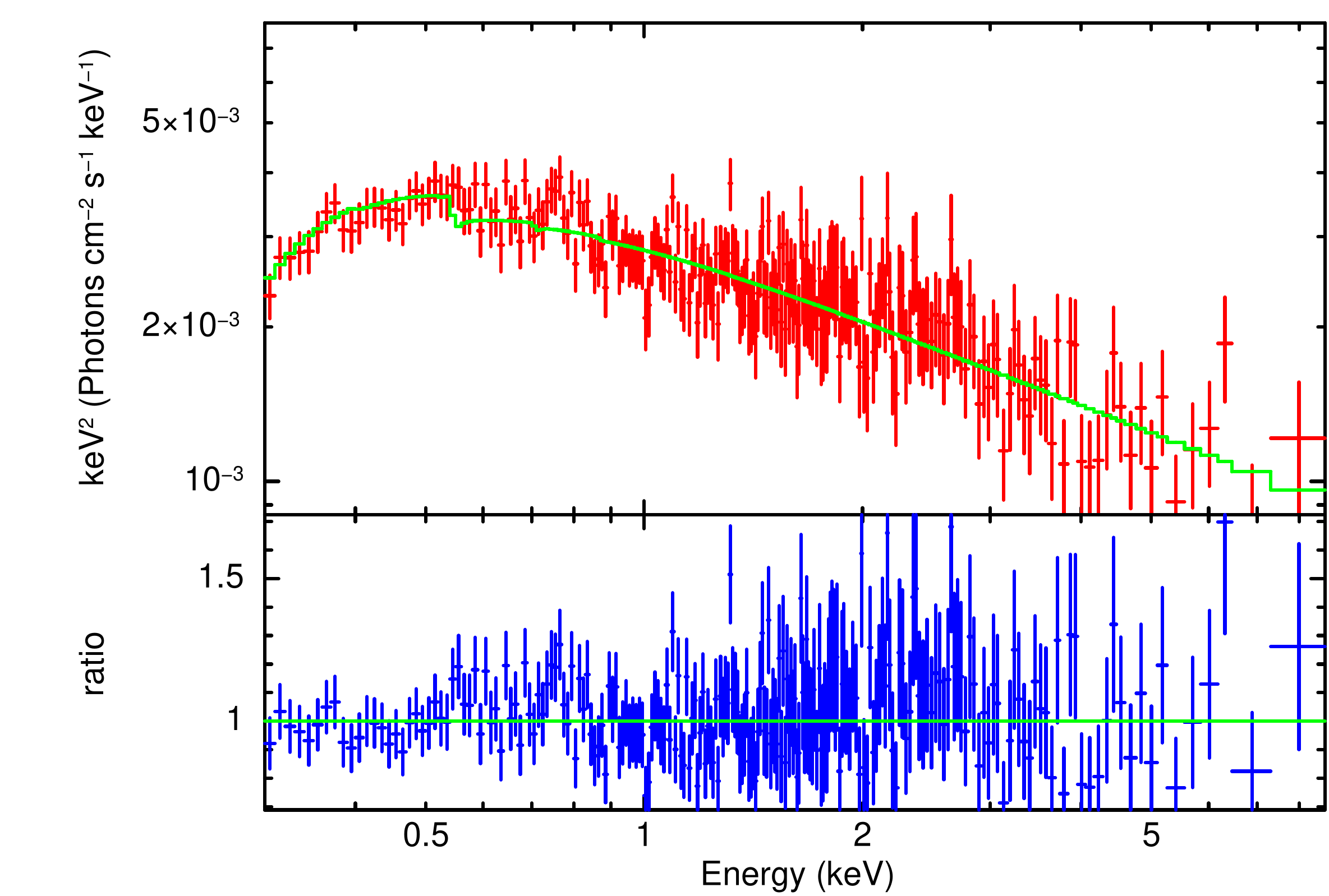}
    \caption{ The \swift-XRT X-ray spectrum for all the three states fitted by powerlaw. Lower panel show the residuals of the fit. }
    \label{fig:xray-sed}
\end{figure}

We also compute flux using the \texttt{cflux} model in XSPEC.
The power-law spectral index can be seen changing from low state to high state. Details of our spectral fit results including the fluxes and spectral index, for all the three states, are presented in Table \ref{tab:table1}.
We observe a clear trend of \enquote{softer-when-brighter} behavior of the source OJ 287 on a longer time-scale. 
The spectrum produced for all the three states is shown in Figure \ref{fig:xray-sed} , which clearly show that the shape of the spectrum is also changing when the source transitions from the low state to the intermediate state and finally to the high state. 
The spectral fitting of all the individual observations since 2015 are done in \citet{Komossa_2020} with a single power law and they report a range of spectral index between 1.6-3.0. Our results are based on shorter time scale from 2019 to 2020 and estimated for three different flux states by modeling with the PL and LP spectral models.

\begin{figure}
    \centering
    \includegraphics[scale=0.45]{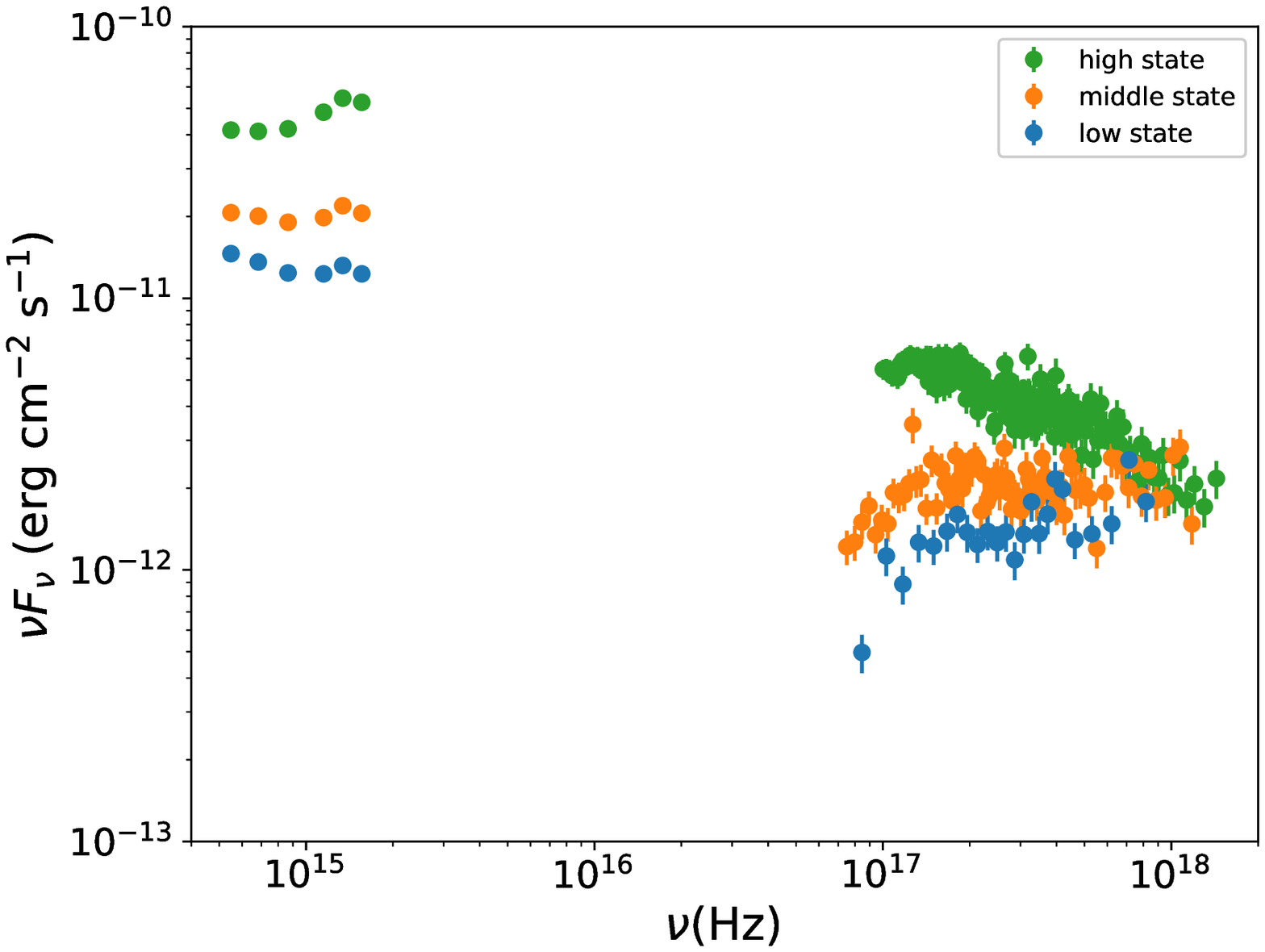}
    \caption{Combined \swift-XRT and UVOT broadband spectrum of all states. A clear spectral change is observed from low state to higher state.}
    \label{fig:xray-uv-sed}
\end{figure}

 We have also extracted the spectrum from the optical-UV band. The combined optical-UV and X-ray spectrum for all the three states is shown in Figure \ref{fig:xray-uv-sed}. Similar to the case in X-ray spectrum, a significant change is noticed here as the source transitions from the low state to the high flux state, which leads to an emergence of a new HBL component and a shift in the synchrotron peak towards higher energy as was previously reported by \citet{Komossa_2020, Komossa2021c}.
As a result, OJ 287, a well-known LBL (\citealt{Nilsson_2018}) blazar, behaves like an HBL type source. A recent study of the kinetic features of radio jet blazars classifies OJ 287 as IBL type blazar (\citealt{Hervet_2016}).
The details about the new HBL component are discussed in section 5.

In Figure \ref{fig:xray-uv-sed}, the shape of the spectrum is flipping around a pivotal energy value corresponding to $\sim$10$^{18}$~Hz as the source changes its states. In the UVOT spectrum, it is found that the spectrum is quite steep during the low state, and then it becomes flat in the intermediate state, and finally, in the high state, it is increasing with energy.  
In the section 5, we have modeled the UVOT and X-ray spectrum for a better understanding of the synchrotron peak during various states.

\subsection{AstroSat, \textit{Swift}-XRT, and NuSTAR  Spectral Analysis}
\begin{figure}
    \centering
    \includegraphics[scale=0.37]{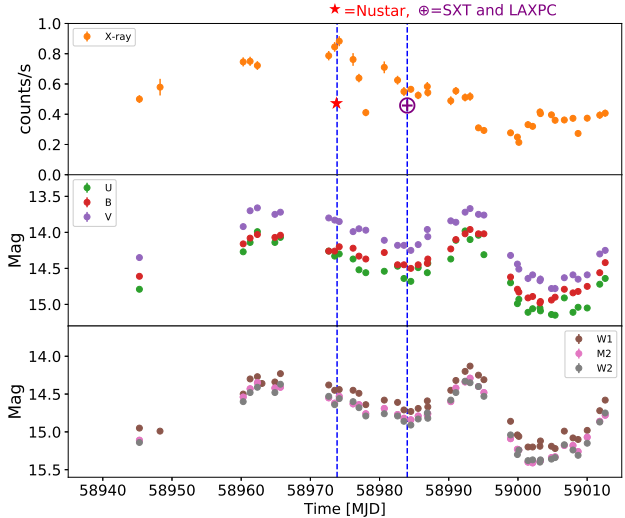}
    \caption{Broadband light curve of 2020 along with the simultaneous observations in other X-ray low and high energy band. Panel 1 shows the \swift-XRT light curve and Panel 2 and 3 represents the \swift-UVOT light curves. Star symbol marked on the panel 1 represents the time of \nus observations and the symbol + sign inside circle represents the time of \asat SXT and LAXPC observations. }
    \label{fig:joint-lc}
\end{figure}

In this section, we present the spectral analysis of simultaneous X-ray observations shown in Figure \ref{fig:joint-lc}.  
 We have used the observation of \nus performed on May 04, 2020 (MJD 58973.86) by \citet{Komossa_2020}, and one observation of \asat-SXT on May 15, 2020 (MJD 58984.0) performed by us. 
Simultaneous \swift observations were also made under the program run by \citet{komossa2017, Komossa_2020}, \citet{2021POBeo.100...29K, universe7080261, Komossa2021c}, we have taken the data from the archive.

\begin{figure}
    \centering
    \includegraphics[scale=0.35,angle=-90]{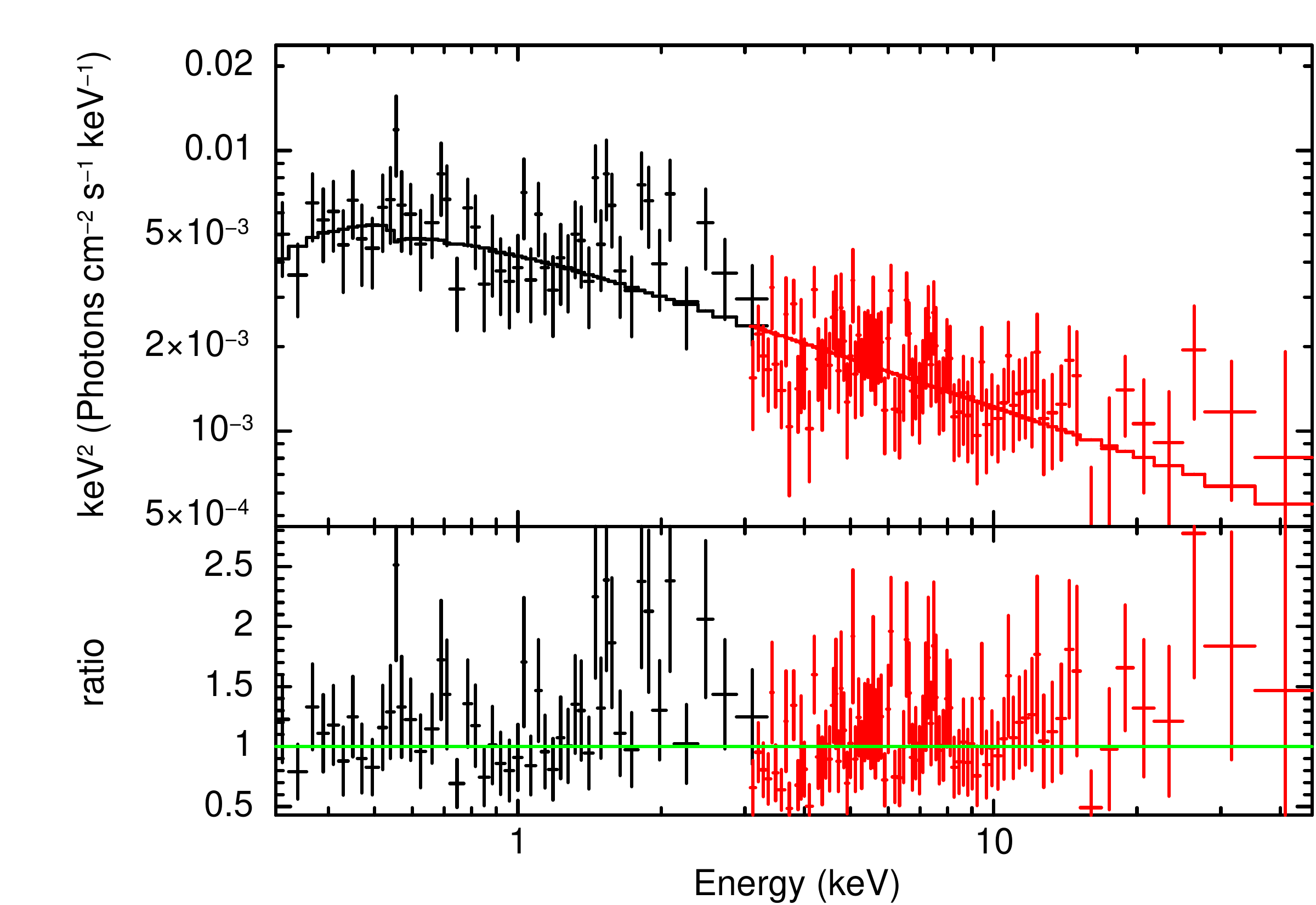}
    \caption{Joint XRT+\nus fitting of the simultaneous observations. Black data points shows the \swift-XRT result and the red data points are the simultaneous \nus results. Lower panel shows the residual of the fit.}
    \label{fig:xrt-nustar}
\end{figure}

In the case of the simultaneous \nus and \textit{Swift}-XRT observations, we have fitted the joint \textit{Swift}-XRT and \nus spectrum to have a clear understanding of the broadband spectrum.
The \textit{Swift}-XRT observation with ObsID:35905053 is simultaneous to the \nus observation, and hence a joint XRT+\nus fit is produced, and a joint spectral fit of XRT+\nus observations is shown in Figure \ref{fig:xrt-nustar}. The joint fitting is performed in \texttt{Xspec} with a power-law spectral model. It is found that a single power-law can explain the total spectrum ranging from 0.3--50.0~keV. The model used for fitting is \texttt{constant*tbabs*powerlaw}. The component \texttt{constant} is added to match the normalization of both the data group and \texttt{tbabs} is used to account for galactic absorption. The best fit power-law index is found to be 2.37$\pm$0.09 and the flux estimated in 0.3-10.0 keV is (2.10$\pm$0.07)$\times$10$^{-11}$ erg cm$^{-2}$ s$^{-1}$ and the flux in 3.0--50.0~keV is found to be (6.34$\pm$0.20)$\times$10$^{-12}$~erg cm$^{-2}$ s$^{-1}$.

\begin{figure}
    \centering
    \includegraphics[scale=0.35,angle=-90]{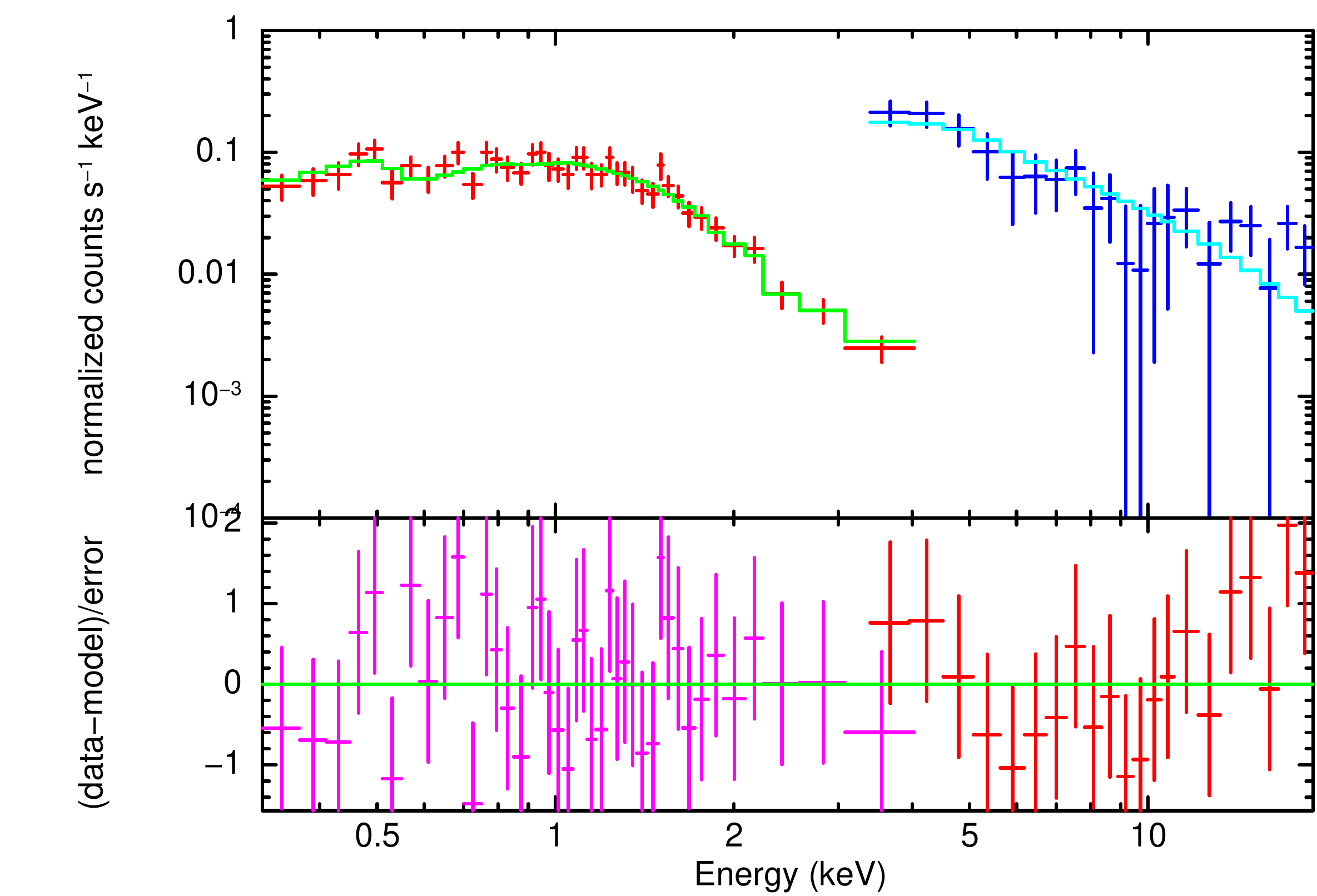}
    \caption{Joint \asat SXT+LAXPC fitting for the simultaneous observations done on May 15, 2020. In upper panel, red  data points represent the SXT data and blue data points are the data from LAXPC instrument. Lower panel shows the residual of the fit.}
    \label{fig:sxt-spec}
\end{figure}

\asat SXT and LAXPC observations are simultaneous as marked in Figure \ref{fig:joint-lc}. The joint SXT+LAXPC spectral fitting is done using the power-law model along with the galactic absorption and the constant factor, same as described in the above section. The spectral fitting plot is shown in Figure \ref{fig:sxt-spec}. The spectral model \texttt{constant*tbabs*powerlaw} is used to model the spectrum. The simultaneous XRT observation corresponding to OBSID 00035905063 was also analyzed separately, and the spectral index was found to be 2.67$\pm$0.14. Comparing this value with the $\Gamma$ = 2.43$\pm$0.09 obtained from the joint SXT+LAXPC spectral fit, suggests that the XRT spectrum is comparatively steeper than the joint SXT+LAXPC spectrum.
The shape of the joint spectra of XRT+\nus and SXT+LAXPC are consistent with each other. 

The best fit parameters of all the simultaneous and quasi-simultaneous spectra are presented in Table \ref{tab:table2}. 
\nus spectrum is also fitted in \citet{Komossa_2020} with powerlaw spectral model and they found the spectral index 2.36$\pm$0.06 and 2.20$\pm$0.20 below 10.0 keV and after the 10.0 keV, respectively. Our joint spectral fitting of XRT+\nus with single powerlaw shows the spectral index 2.37$\pm$0.09 which is consistent with theirs below 10.0 keV spectrum. We have also did the joint spectral fitting of SXT+LAXPC with single powerlaw and the the joint spectral index estimated to be 2.43$\pm$0.09 which shows consistency with our XRT+\nus joint fit. Here the SXT and LAXPC joint fit is important since the spectrum is highly simultaneous and the exposure time is $\sim$ 30 ks. Due to the higher exposure time \asat-SXT spectrum is much better than the \swift-XRT spectrum used in the joint fitting (Figure \ref{fig:xrt-nustar} \& \ref{fig:sxt-spec}).

\section{Modeling the Synchrotron peak: HBL like component}

The synchrotron peak in blazars can be constrained by the near-infrared and optical-UV emission along with the soft X-ray emission. Previously, a log-parabolic function \citep{Massaro2004, Kapanadze_2018} and a log-cubic function \citep{Raiteri2013} have been used to model the NIR, optical-UV, and X-ray spectrum. In our work, we have used a different approach, where we modeled the optical-UV and X-ray data points with the synchrotron process using the publicly available code GAMERA; modeling details can be found in \citet{Prince_2021}.

\begin{figure}
    \centering
    \includegraphics[scale=0.35]{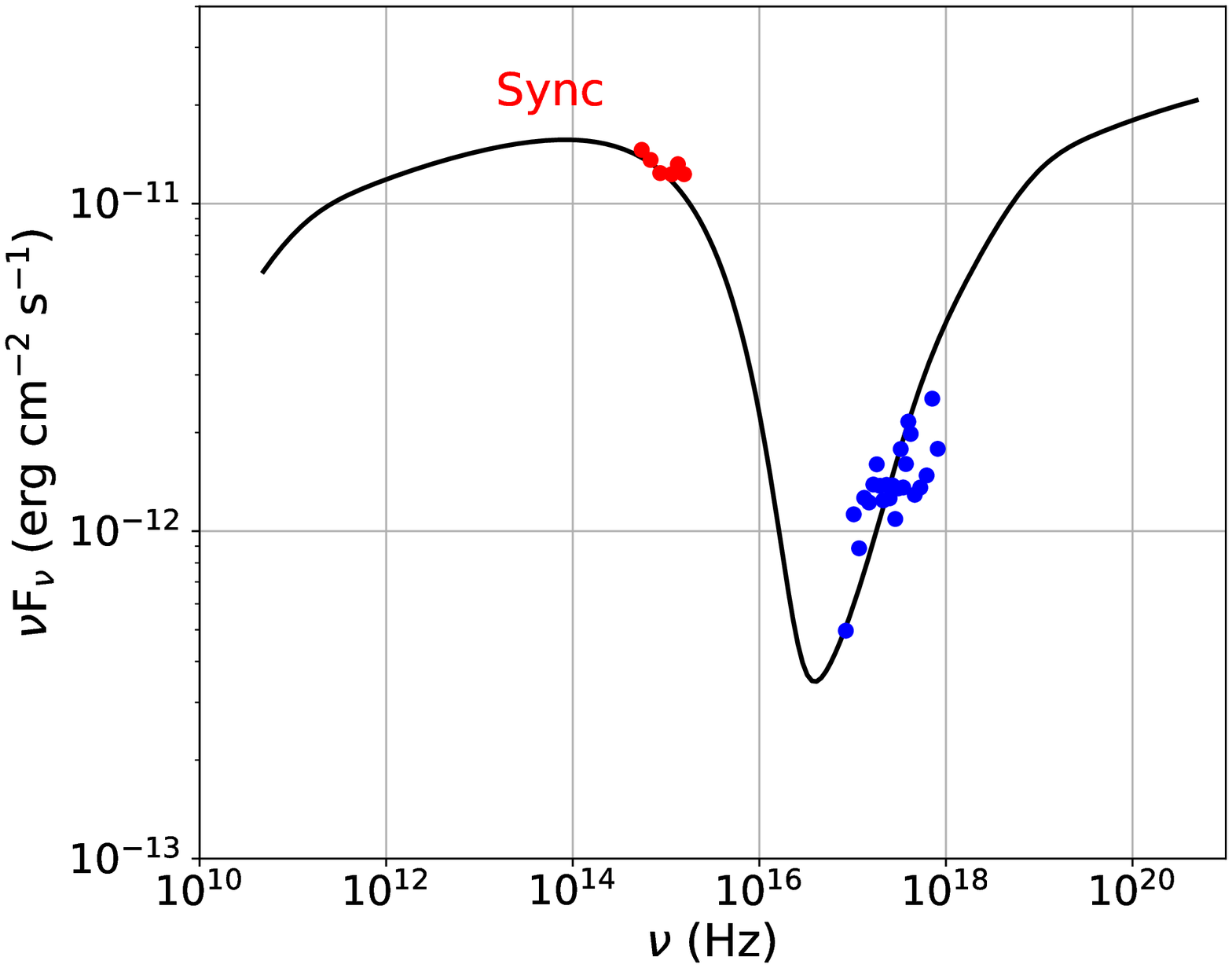}
    \includegraphics[scale=0.35]{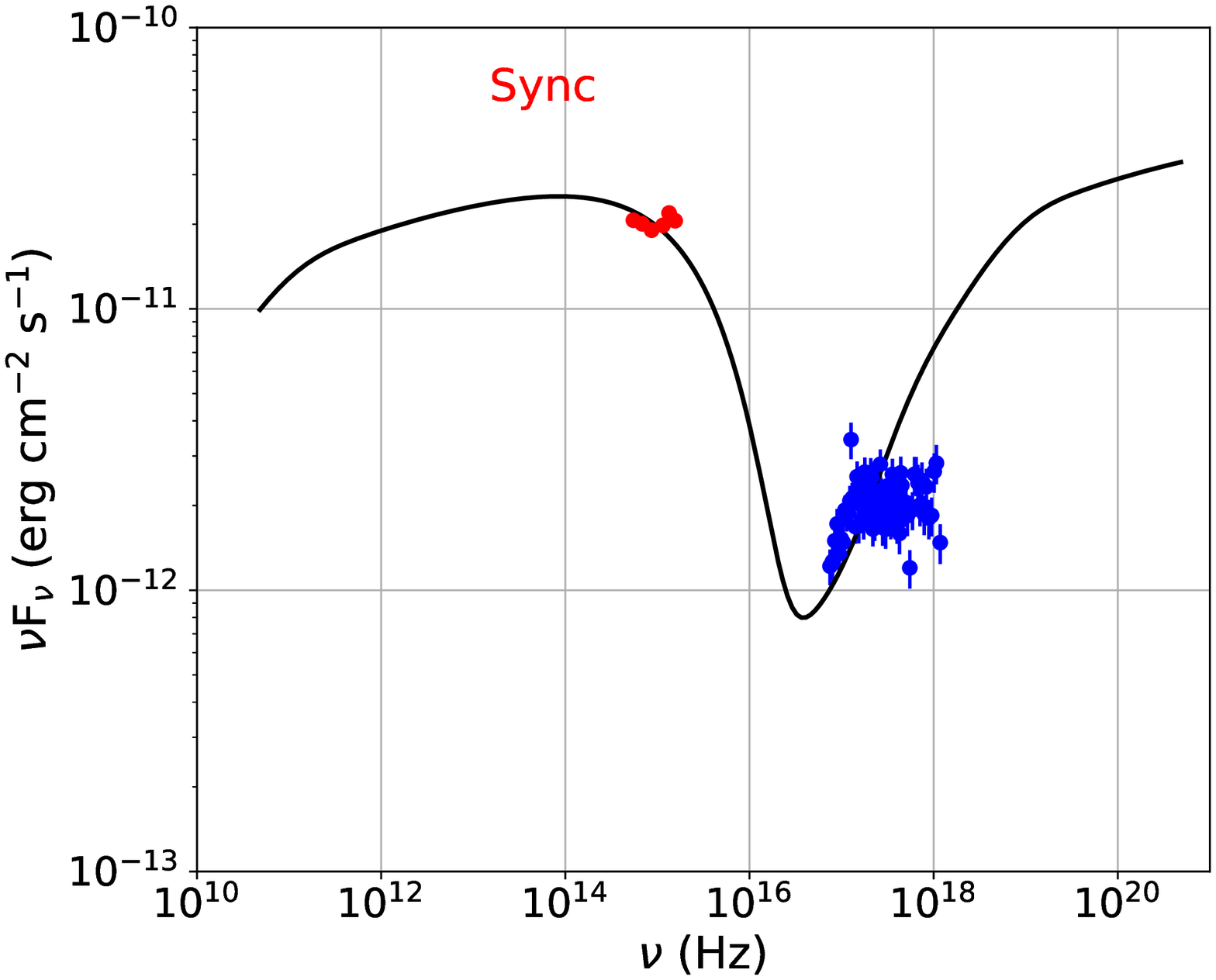}
    \includegraphics[scale=0.35]{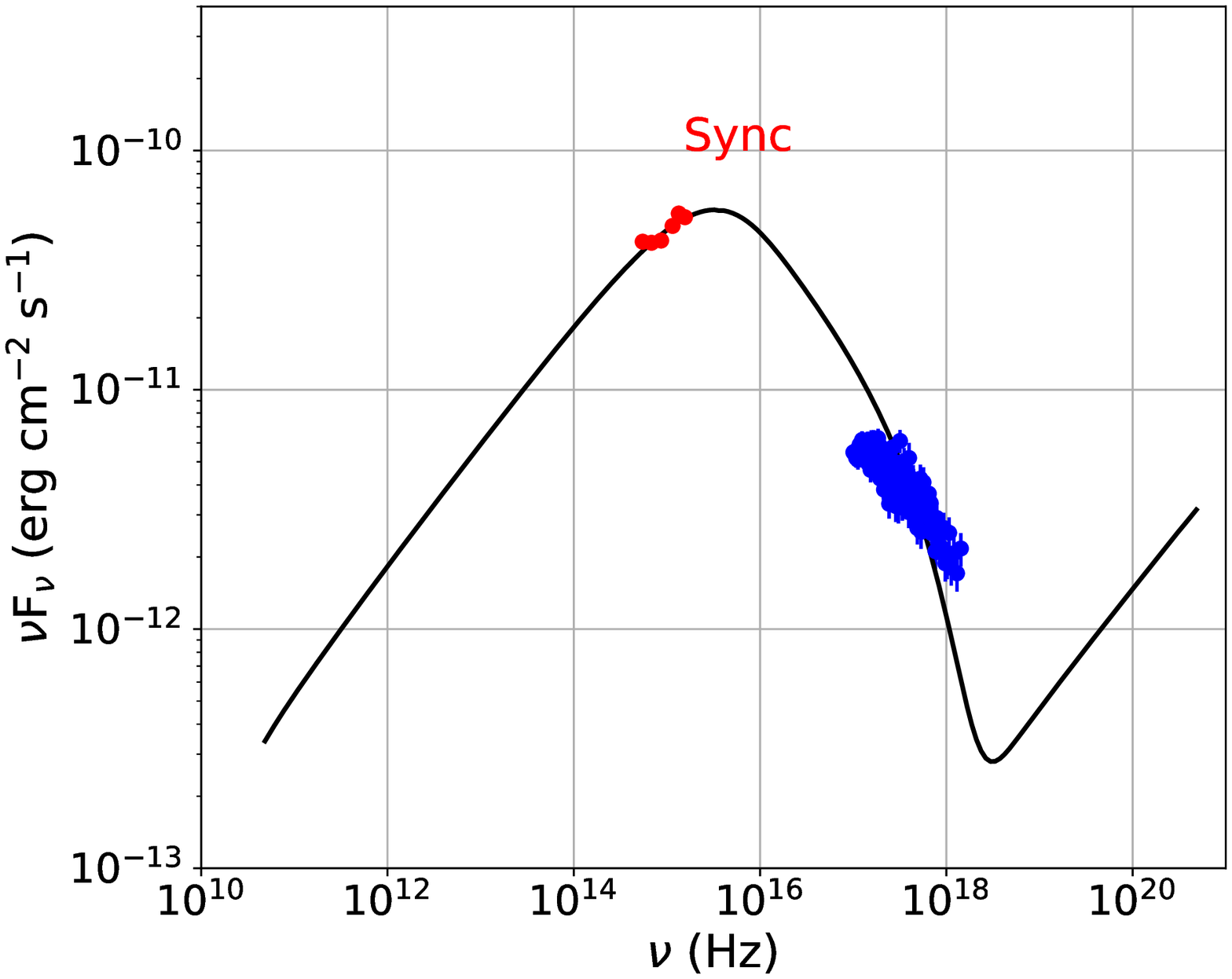}
    \caption{Broadband SED fit during the low, middle, and high state. The data points are from \swift-XRT/UVOT and the synchrotron peak is fitted with code GAMERA.}
    \label{fig:sync-shift}
\end{figure}

We found that during the low state of the source, the synchrotron peak is observed at around $\sim$10$^{14}$~Hz, which is the synchrotron peak assigned for the LBL type source  (\citealt{Padovani1995}, \citealt{abdo_2010}). During the intermediate state, the X-ray spectrum does not change much, however a marginal shift of the synchrotron peak towards the higher energy is observed because of change observed in UVOT spectrum (Figure \ref{fig:sync-shift}).
During the high flux state, the shape of the UVOT and X-ray spectrum both have changed significantly compared to the low flux state, suggesting the emergence of a new HBL component in this source. The fitting of the parameters using GAMERA shows that a higher magnetic field ($\sim$6.7~G) is required in high flux state. However, in the low and intermediate state the magnetic field is found to be $\sim$4.1~G. The LP electron distribution is provided to GAMERA to model the synchrotron peak. The spectral index, $\alpha$ is 1.8 for low and intermediate state and a steeper spectral index of  2.65 is required for the high state.
Our modeling also suggests the need for higher energy electrons during high state compared to low and intermediate state. The values of the parameters obtained in this work are consistent with the parameters estimated in  Prince et al. 2021 (under review).
Modeling of the synchrotron emission shows that the synchrotron peak is  shifted towards the higher energy, and the current peak location is around 10$^{16}$~Hz, which is the ideal case for the HBL type BL Lac source (\citealt{Padovani1995}, \citealt{abdo_2010}). The shift is two orders of magnitude in frequency which is very rarely seen in a single blazar. The results suggest that during the flaring period, the source exhibits very different spectral properties compared to when it is in a normal state. The result is shown in Figure \ref{fig:sync-shift}. In OJ 287, this behavior was also seen before during the flare of 2016-2017 (\citealt{kushwaha_2018b}) where the synchrotron peak was shifted towards the higher energy. The first appearance of the HBL component was seen in a flare of 2017, the study of \citet{Kushwaha_2020} discussed the disappearance of the HBL component in 2018. Again the HBL component appeared in the flare of 2020 as mentioned by \citet{Komossa_2020}, \citet{Kushwaha_2020}, which has also been found in the present study.
This finding suggests that the source is highly variable and changes its behavior during high flux state and fluctuates between LBL and HBL type source, and has an identity crisis. The blazar classification into FSRQ and BL Lac, and further the classification of BL Lac objects into LBL, IBL, and HBL is based on the fixed location of the synchrotron peak (\citealt{abdo_2010}). Therefore, the change in the synchrotron peak is not very common, and if it does occur, it challenges the understanding of the underlying physical mechanisms in these types of sources. Blazar OJ 287 is a well accepted binary black hole system, and the significant change in the optical-UV and X-ray spectral state could be associated with the disk-impact scenario. However, in the binary black hole model, the black hole-disk impact was expected in July-September 2019, and because of the sun's constraint it could not be observed using \swift  but was observed by the \spitzer  (\citealt{Laine_2020}). Considering the spectral change is associated with the disk-impact scenario, the appearance of new HBL components in April-May 2020 is still puzzling and needs more future study to draw any reliable conclusion.
One of the possible explanation for the new HBL component could be the increase in the accretion rate (\citealt{Komossa_2020}) post BH-disk impact and consequently it could trigger the flare produced in the jet after a few months. The time delay observed around nine months between the impact and the time of flare is not easy to understand, since various physical conditions could be responsible for that (disk/corona  properties  and  geometry,  magnetic field geometry, shock formation in the jet).

In our first paper, we have modeled the broadband SED of 2020 flare along with a low state, and we found that one emission zone is sufficient to explain the total SED (\citealt{Prince-2021}). 

Recent study by \citet{Komossa2021c} has discussed the long term X-ray spectral behavior of the source. They analysed the \xmmn and \swift data from 2005--2020, and found that the X-ray spectrum is highly variable. They also mentioned that the X-ray spectrum covers all the behaviors seen in blazars from significant flat to very steep.  

\begin{table*}
    \centering
    \begin{tabular}{c|ccc|cccc|cc}
    \hline
    \noalign{\smallskip}
 Observation ID & PL & &&LP& & &&F-test & p-value \\
 &F$_{0.3-10 keV}$ & $\Gamma$& ${\chi}^2$ (dof)& F$_{0.3-10 keV}$&$\alpha$& $\beta$ &${\chi}^2$(dof)  & &  \\
  \noalign{\smallskip}
  \hline \noalign{\smallskip}
 low state&5.15$\pm$0.15 & 1.89$\pm$0.07 & 85.30(86)& 4.98$\pm$0.20&1.86$\pm$0.09 & 0.12$\pm$0.21 & 84.46(85) & 0.84 & 0.36 \\
   \noalign{\smallskip}
  \hline \noalign{\smallskip}
 intermediate state&7.14$\pm$0.08 & 2.15$\pm$0.04 & 166.40(145)& 7.22$\pm$0.10 &2.16$\pm$0.05 & 0.04$\pm$0.11 & 166.01(144) & 0.34 & 0.56 \\
 \noalign{\smallskip} \hline \noalign{\smallskip}
  high state & 14.72$\pm$0.09 &2.56$\pm$0.02 & 206.14(227)& 14.60$\pm$0.12 &2.55$\pm$0.02 & 0.04$\pm$0.07 & 205.04(226)& 1.21 & 0.27 \\
\noalign{\smallskip} 
  \hline 
    \end{tabular}
    \caption{Modeled parameters for the X-ray spectrum during low state, intermediate state, as well as of high state. The fluxes are in unit of 10$^{-12}$ erg cm$^{-2}$ s$^{-1}$.}
    \label{tab:table1}
\end{table*}

\begin{table}
    \centering
    \begin{tabular}{c|ccc|}
    \hline
    \noalign{\smallskip}
 Instruments & PL & & \\
and Dates &F$_{0.3-10 keV}$ & $\Gamma$& ${\chi}^2$ (dof)\\ 
& (F$_{3.0-50.0 keV}$) & & \\ 
  \noalign{\smallskip}
  \hline \noalign{\smallskip}
 SXT + LAXPC &0.91$\pm$0.04 & 2.43$\pm$0.09 & 40.25(57) \\
            & (0.96$^-_-$) & & \\
   \noalign{\smallskip}
  \hline \noalign{\smallskip}
 \noalign{\smallskip} \hline \noalign{\smallskip}
  XRT + \nus & 2.10$\pm$0.17 &2.37$\pm$0.09 & 119.22(134)\\
 & (0.63$\pm$0.04) & & \\
\noalign{\smallskip} 
  \hline 
    \end{tabular}
    \caption{Modeled parameters for the simultaneous X-ray spectrum from XRT, SXT, \nus, and LAXPC during various occasions. The fluxes are in unit of 10$^{-11}$ erg cm$^{-2}$ s$^{-1}$.}
    \label{tab:table2}
\end{table}
\section{Color-magnitude diagram}
Simultaneous monitoring of OJ 287 in various optical (U, B, V) bands allow us to study the color variation of the source. The color-magnitude diagram helps in understanding the origin of different physical processes responsible for the flux variations in blazars.
We fit the plots of colour indices (CIs) vs. magnitude (M) with
straight lines (i.e CI = m*M +c) using which we estimated the fit values of the
slope, m, the Spearman correlation coefficient r
along with the corresponding null hypothesis probability, p which
are summarized in Table \ref{tab:color-mag}. A positive slope (when the p value is $<$ 0.05) indicates significant positive correlation between CI and blazar magnitude, which in turn points towards a bluer when brighter (BWB) or redder when fainter
trend (H.E.S.S. Collaboration et al. 2014) in the target, while a negative slope
implies the opposite i.e. redder when brighter (RWB) trend \citep{2019MNRAS.488.4093A}.

\begin{figure*}
    \centering
    \includegraphics[scale=0.55]{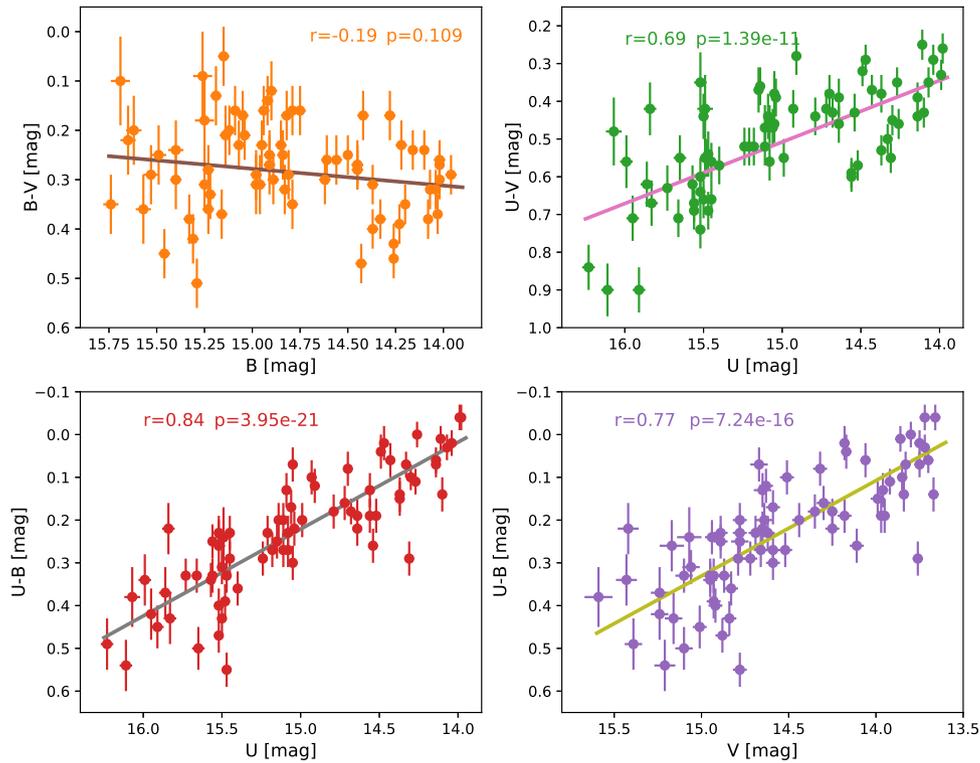}
    \caption{Colour-magnitude diagram during the 2020 flare.}
    \label{fig:col-mag}
\end{figure*}

We have plotted the four color-magnitude diagram defined among B-V vs. B, U-V vs. U, U-B vs. U, and U-B vs. V, which are shown in Figure \ref{fig:col-mag}. The estimated values of $r$ and $p$ for all the cases are mentioned in each plot. In three cases, we have seen a very strong correlation with a correlation coefficient of 0.69, 0.84, and 0.77, respectively, suggesting a \enquote{bluer-when-brighter} chromatism (Figure \ref{fig:col-mag}). In the case of B-V vs. B, we did not find any strong correlation; however, the slope suggests RWB chromatism but could not be confirmed. 
\par

Previous study on the color-magnitude diagram have shown various trends, including the BWB and RWB. BWB chromatism has been previously reported in the B and V bands between 1973--1976 \citep{Carini_1992}, in the R and V bands between 1993--1997 \citep{Dai_2011}, in the V and J bands (weaker correlation) between 2008--2010 \citep{Ikejiri_2011} and in the Ringo3 bands in 2015 \citep{jermak_2016}.
A recent study by \citet{Wierzcholska_2015} found weak or no correlation on longer timescales in the B and R bands, while they claim the BWB chromatism at shorter isolated intervals in the long-term light curve. Another study by \citet{Siejkowski_2017} did not find correlation in any combination of color-magnitude diagrams. 
\citet{Gupta_2019} found
BWB chromatism in V-R and R-I colors with respect to R band magnitude.  However, their Pearson's correlation coefficient is below 50\%.

A clear BWB and RWB and occasional lack of correlation in the color-magnitude diagram of OJ 287 indicate a complex nature of the physical process responsible for the optical emission in this blazar. 

Further, we have also estimated the average optical spectral index following the expression from \citet{Wierzcholska_2015},
\begin{equation}
\langle \alpha_{BR} \rangle = \frac{0.4 \langle B-R \rangle} {log(\nu_B/\nu_R)}   
\end{equation}
Where $\langle B-R \rangle$ is the average color from any two bands, and the $\nu_B$, $\nu_R$ are the effective frequency of the corresponding B, and R bands.
The optical spectral index estimated for the \textit{Swift} U, B, V bands vary between 0.89 to 1.15, suggesting a significantly harder spectrum in optical bands. The optical spectral index is shown in Table \ref{tab:color-mag}. A significant spectral hardening in the optical emission is also seen in \citet{Kushwaha_2020}. During the same period, a significant spectral softening is observed in Figure \ref{fig:HR}, suggesting an anti-correlation with optical spectral index.  
 \citet{Wierzcholska_2015} find that OJ 287 has a significantly steep optical spectrum, between, $\alpha_{BR}$ $\sim$2-3. In our study, the anti-correlation between optical and X-ray spectral index might suggests they are produced through different processes. Many more optical and X-ray study is required to come-up with any concrete conclusions.

\begin{table}
    \centering
    \begin{tabular}{c|cccc}
    Color & m   & r  & p-value & $\langle \alpha_{BR} \rangle$  \\
    \hline
    B-V & -0.03 & -0.19 & 0.109 & 1.15$\pm$0.22\\
U-V & 0.16  & 0.69  & 1.39e-11& 1.01$\pm$0.10 \\
    U-B & 0.20  & 0.84  & 3.95e-21 & 0.89$\pm$0.17 \\
    U-B * & 0.22  & 0.77  & 7.24e-16& - \\
  \hline
    \end{tabular}
    \caption{Calor magnitude variations and their parameters. m = slope, r = Pearson's correlation, p-value = Null hypothesis probability, U-B * = color variation with respect to V magnitude. The last column show the average optical spectral index.}
    \label{tab:color-mag}
\end{table}

\section{Summary and Conclusions}

In this work, we present a joint X-ray and UV-optical study of the interesting blazar OJ 287 during its flaring activity between 2019--2020 using \asat, \nus and \swift observations. The study includes the data from \textit{AstroSat}-SXT and LAXPC instruments. We categorized the long term \swift light curve as low state, intermediate state, and high state (Figure \ref{fig:mwl-lc}). The flare of 2020, which is categorized as a soft X-ray flare, shows a \enquote{softer-when-brighter} behavior. Joint spectral fitting of simultaneous observations of \textit{Swift}-XRT and \nus, and the \textit{AstroSat}-SXT and LAXPC indicate consistent power-law spectral index values of 2.37$\pm$0.09 and 2.43$\pm$0.09 in the 0.3--50.0~keV and 0.3--20.0~keV bands, respectively. Since \asat-SXT has a longer exposure compared to \swift-XRT, it has been very useful in generating sufficient SNR to better constrain the source spectral properties. Our spectral study indicates  a shift in the synchrotron peak of the source towards the higher energy which suggests an emergence of HBL components during the high flux state. We have modeled the synchrotron peak with the physical model implemented in publicly available code GAMERA. Our modeling suggests that a higher value of magnetic field is required to explain the high state compared to low state in leptonic scenario. Results from our study of the color-magnitude behavior of the source during this bright state show significant correlations in all bands, except in the B-V vs. B bands, suggesting a strong \enquote{bluer-when-brighter} chromatism. A significant harder spectrum is observed in optical which is anti-correlated with the X-ray emission.  The different behaviors of OJ 287 observed during various flux states suggest a complex nature of the source. Much more detailed optical and X-ray study would be required for better understanding of this proposed binary black hole system. 

\section*{Acknowledgements}
The project was partially supported by the Polish Funding Agency National Science Centre, project 2017/26/A/ST9/00756 (MAESTRO 9), and MNiSW grant DIR/WK/2018/12.

\section*{Data Availability}
The data and softwares used in this research are available at NASA’s HEASARC
webpages with the links given in the manuscript.

\vspace{5mm}

\bibliographystyle{mnras}
\bibliography{reference-list.bib}



\end{document}